\pdfoutput=1

\documentclass[10pt,aps,prx, twocolumn,superscriptaddress,notitlepage,longbibliography]{revtex4-2}

\usepackage{setspace}

\usepackage{booktabs}

\usepackage{tabularx}
\usepackage{amsmath}
\usepackage{amssymb}
\usepackage{graphicx}
\usepackage{bm}

\usepackage{txfonts}

\hbadness=\maxdimen
\vbadness=\maxdimen
\def\be{\begin{equation}}
\def\ee{\end{equation}}
\def\bea{\begin{eqnarray}}
\def\eea{\end{eqnarray}}
\def\ba{\begin{array}}
\def\ea{\end{array}}

\usepackage{xcolor}
\usepackage[colorlinks=true,allcolors=blue]{hyperref}

\allowdisplaybreaks

\begin{document}

\title{Criticality enhances the reinforcement of disordered networks by rigid inclusions}% 

\author{Jordan L.\ Shivers}
\affiliation{Department of Chemical and Biomolecular Engineering, Rice University, Houston, TX 77005, USA}
\affiliation{Center for Theoretical Biological Physics, Rice University, Houston, TX 77005, USA}
\affiliation{James Franck Institute, University of Chicago, Chicago, IL 60637, USA}
\affiliation{Department of Chemistry, University of Chicago, Chicago, IL 60637, USA}
\author{Jingchen Feng}
\affiliation{Center for Theoretical Biological Physics, Rice University, Houston, TX 77005, USA}
\author{Fred C.\ MacKintosh}
\affiliation{Department of Chemical and Biomolecular Engineering, Rice University, Houston, TX 77005, USA}
\affiliation{Center for Theoretical Biological Physics, Rice University, Houston, TX 77005, USA}
\affiliation{Department of Chemistry, Rice University, Houston, TX 77005, USA} 
\affiliation{Department of Physics \& Astronomy, Rice University, Houston, TX 77005, USA}

\begin{abstract}
The mechanical properties of biological materials are spatially heterogeneous. Typical tissues are made up of a spanning fibrous extracellular matrix in which various inclusions, such as living cells, are embedded.  While the influence of embedded inclusions on the stiffness of common elastic materials such as rubber has been studied for decades and can be understood in terms of the volume fraction and shape of inclusions, the same is not true for disordered filamentous and fibrous networks. Recent work has shown that, in isolation, such networks exhibit unusual viscoelastic behavior indicative of an underlying mechanical phase transition controlled by network connectivity and strain. How this behavior is modified when inclusions are present is unclear. Here, we present a theoretical and computational study of the influence of rigid inclusions on the mechanics of disordered elastic networks near the connectivity-controlled central force rigidity transition. Combining scaling theory and coarse-grained simulations, we predict and confirm an anomalously strong dependence of the composite stiffness on inclusion volume fraction, beyond that seen in ordinary composites.  This stiffening exceeds the well-established volume fraction-dependent stiffening expected in conventional composites, e.g., as an elastic analogue of the classic volume fraction dependent increase in the viscosity of liquids first identified by Einstein. We show that this enhancement is a consequence of the interplay between inter-particle spacing and an emergent correlation length, leading to an effective finite-size scaling imposed by the presence of inclusions. We outline the expected scaling of the linear shear modulus and strain fluctuations with the inclusion volume fraction and network connectivity, confirm these predictions in simulations, and discuss potential experimental tests and implications for our predictions in real systems.
\end{abstract}

\maketitle

\vspace{1em}

\section{Introduction}

Composite materials, composed of two or more distinct material components, can exhibit surprising and often useful emergent properties not seen in pure samples of the individual components. Over the last century, there has been a rich scientific and engineering effort to understand how the mechanical response of viscoelastic materials can be modified by the incorporation of mechanical inclusions.  Perhaps the most prolific class of such materials are rubber-based composites, which see widespread use in industrial and consumer goods. The addition of even small concentrations of stiff particulate inclusions to rubber can produce dramatic increases in stiffness. Initial efforts to theoretically explain this effect were limited to the very dilute regime, where a linear dependence of overall stiffness on inclusion volume fraction is observed \cite{smallwood_limiting_1944, guth_theory_1945}, and were based on Einstein's early work on the viscosity of dilute suspensions of spherical particles \cite{einstein_neue_1906,einstein_berichtigung_1911}. A significant body of subsequent work has enabled the prediction of properties at larger volume fractions \cite{torquato_random_2002} up to and, for nonrigid particles, even beyond the jamming threshold \cite{zhao_elasticity-controlled_2024}.

Less is understood about biological composites, such as living tissues, in which one or more of the material components is typically a disordered biopolymer network.  It is known experimentally that inclusions can strongly influence the mechanical response of biopolymer-based composites, both providing mechanical reinforcement evident in the linear elastic properties \cite{song_strain_2023} and even producing qualitatively unexpected nonlinear phenomena, such as compression-driven stiffening behavior \cite{gandikota_loops_2020,shivers_compression_2020,song_strain_2023} seen in various tissues \cite{pogoda_compression_2014,perepelyuk_normal_2016,van_oosten_emergence_2019}. Notably, inclusions in biological composites are often not inert particles, but are instead active entities that can exert stress on and even remodel the surrounding material \cite{parker_how_2020}. For example, contractile cells can induce long-range stiffening, reorientation and directed force transmission in the extracellular matrix \cite{vader_strain-induced_2009, jansen_cells_2013, wang_long-range_2014, han_cell_2020,muntz_role_2022, zhang_enhanced_2024}. In blood clots, contracting platelets can significantly stiffen their surrounding fibrin networks \cite{shah_strain_1997,lam_mechanics_2011,pathare_fibrin_2021, qiu_biophysics_2019,zakharov_clots_2024}. 
In the extracellular matrix, individual cells \cite{friedl_collective_2009,bonnans_remodelling_2014, abhilash_remodeling_2014} and larger cellular aggregates such as tumors \cite{provenzano_cell_2015,naylor_micromechanical_2022, zhang_enhanced_2024} can also structurally remodel their surroundings, producing dramatic changes in the local spatial organization and stiffness of the matrix. Simultaneously, the stiffness and spatial organization of the extracellular matrix play a crucial role in controlling various cellular processes \cite{levental_soft_2007}, such as differentiation \cite{engler_matrix_2006, chaudhuri_hydrogels_2016, chaudhuri_effects_2020} and cell migration \cite{gardel_mechanical_2010,lu_extracellular_2012, koch_3d_2012,paul_cancer_2017,kai_extracellular_2019}. Given the relevance of these phenomena to important biological processes and their potential utility in developing bio-inspired tunable materials, improving our understanding of how inclusions can modulate the mechanics of biopolymer networks remains an important challenge.

Here, we consider the elasticity of composites in which the interstitial matrix is a disordered elastic network, modeled as a bond-bending spring network \cite{feng_percolation_1984, broedersz_modeling_2014}. Specifically, we consider networks in which the average connectivity or coordination $z$ of nodes is in the vicinity of the central force isostatic point $z_c = 2d$, described originally by Maxwell \cite{maxwell_calculation_1864}. In such systems, the distance $\Delta z = z - z_c$ from the critical point governs the elastic response and strain fluctuations or nonaffinity \cite{wyart_elasticity_2008}, which are characterized by a diverging correlation length $\xi \sim \xi_{\pm}|\Delta z | ^{-\nu}$, where $\nu$ is a correlation length exponent \cite{broedersz_criticality_2011}. Without bending interactions, such networks are rigid (for large system size) only if $z \ge z_c$. In the weak bending limit, the mechanical response of these networks is \textit{bending dominated} (soft) for $z < z_c$ and \textit{stretching-dominated} (stiff) for $z \ge z_c$. Consistent with conventional thermodynamic critical phenomena \cite{wilson_problems_1979,binder_finite_1987, cardy_finite-size_2014} and with observations of the behavior of granular systems near the jamming critical point \cite{goodrich_scaling_2016,goodrich_finite-size_2012}, one can observe pronounced finite size effects in the scaling behavior of these networks near the critical point \cite{broedersz_criticality_2011, shivers_scaling_2019}. These finite size effects are a consequence of the limitation that a finite system size imposes on the divergence of the correlation length. Qualitatively, one finds that in the vicinity of the critical point, a reduction in the system size leads to an increase in stiffness and a decrease in nonaffinity, both of which are well captured by finite size scaling theory \cite{broedersz_criticality_2011}. In this paper, we show that when such a network is combined with rigid inclusions to construct a composite, the typical inter-inclusion length scale provides an effective upper bound on the network correlation length, and thus plays the role of system size, such that increasing the inclusion volume fraction actually stiffens the interstitial matrix. This behavior would not be observed if the matrix were a conventional elastic material. Thus, we expect that the stiffening of a critical network-based composite upon the addition of rigid inclusions reflects two cooperative stiffening effects: the standard volume fraction-dependent stiffening effect observed in ordinary elastic materials and the dependence of the matrix stiffness on the volume fraction-dependent inter-inclusion length scale. 

\begin{figure}[ht]
    \centering
    \includegraphics[width=1\columnwidth]{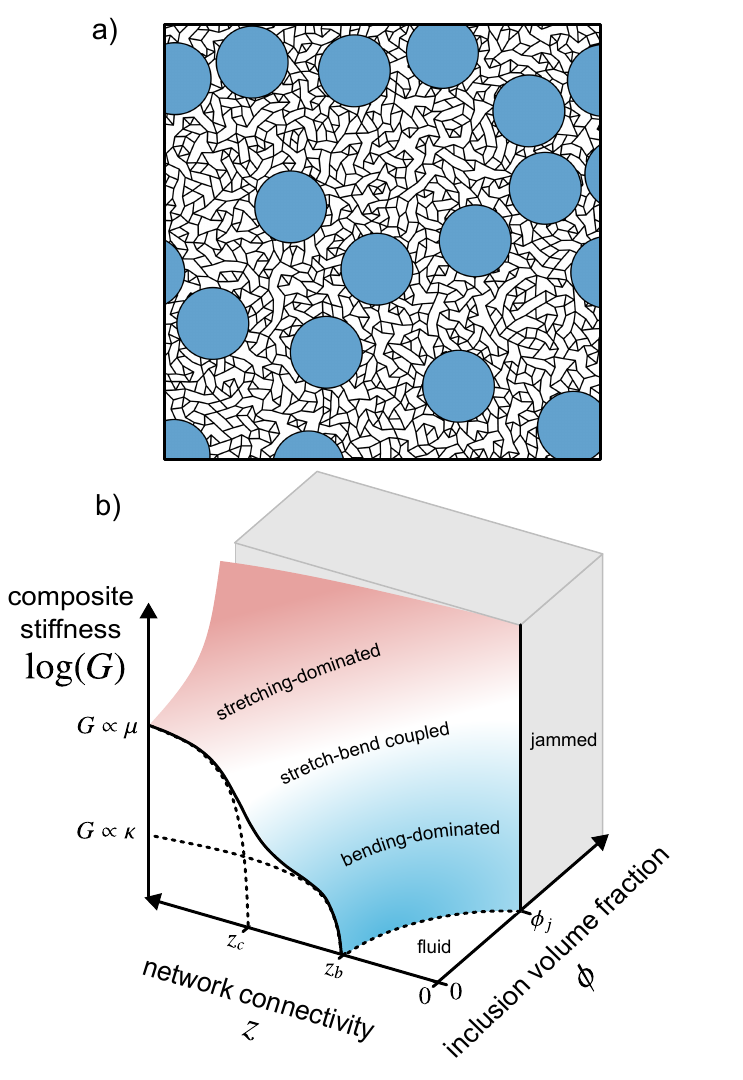}
    \caption{\label{fig:3D_schem}
      \textbf{ Rigid inclusions give rise to dramatic changes in the mechanics of disordered bond-bending networks.} (a) A portion of a simulated two-dimensional disordered network/rigid particle composite. The interstitial matrix is a bond-bending network with average node connectivity $z$ (here, $z = 3.5$), in which springs have linear 1D modulus $\mu$ and pairs of adjacent springs have harmonic bending interactions with stiffness $\kappa$. Rigid disk-shaped inclusions of uniform radius $R$ are randomly placed throughout the network with total volume fraction $\phi$ and are permanentally connected to the surrounding network. Inclusions are rigid but free to translate and rotate, subject to the forces exerted by the attached springs. (b) 
      Schematic representation of the composite stiffness $G$ for disordered elastic networks with connectivity $z$ and rigid inclusion volume fraction $\phi$, for networks with weak bending interactions ($\propto \kappa$) and stiff stretching interactions ($\propto \mu$).}
  \end{figure}

We develop a scaling theory for both of these effects and validate the resulting predictions in simulations of two-dimensional composites. Remarkably, we find that the presence of inclusions can give rise to macroscopically stretch-bend-coupled and even fully stretching-dominated mechanics in otherwise bending-dominated disordered networks. In particular, we show that increasing the inclusion volume fraction leads to a reduction in the critical connectivity, shifting the phase boundary in a manner consistent with conventional finite size scaling \cite{binder_finite_1987}. In the strictly $\kappa\to 0$ limit, the phase boundary demarcates the transition between bending-dominated and stretching-dominated mechanics. However, for finite $\kappa$, the phase boundary is surrounded by a stretch-bend coupled regime \cite{broedersz_criticality_2011}. The proposed behavior is sketched in. Fig \ref{fig:3D_schem}a. As our phase diagrams show, the extent of this regime also shifts to lower values of $z$ as the inclusion volume fraction is increased. We observe this expansion of the critical regime to lower $z$ values both in our measurements of $G$ and in the nonaffine fluctuations.

The paper is structured as follows. In Section \ref{sec:background}, we provide additional background information, before outlining our scaling theory in Section \ref{sec:scalingtheory}. In Section \ref{sec:ordinary}, we briefly review the volume fraction-dependent stiffening observed in ordinary composites. In Section \ref{sec:lengthscaledependent}, we derive the typical inter-inclusion length scale $\xi_p(r,\phi)$. In Section \ref{sec:scaling}, we derive scaling predictions for the dependence of the shear modulus $G$ on the inclusion volume fraction and network connectivity. In Section \ref{sec:simulations}, we describe our approach for simulating two-dimensional networks with embedded inclusions. Finally, in Section \ref{sec:results}, we present and discuss our results.

\section{Background} \label{sec:background}

Biopolymer networks act as viscoelastic scaffolds for cells and tissues and are largely responsible for their remarkable resilience and highly responsive mechanical behavior \cite{kasza_cell_2007,burla_mechanical_2019}.  In recent decades, significant progress has been made toward understanding the physics of biopolymer networks through a rich combination of experiments and modeling \cite{mackintosh_elasticity_1995,kroy_force-extension_1996,satcher_theoretical_1996,shah_strain_1997,weisel_mechanical_2004,gardel_elastic_2004,storm_nonlinear_2005,janmey_fibrin_2009,broedersz_modeling_2014,pritchard_mechanics_2014,jansen_role_2018}. In particular, extensive work has focused on characterizing the static and dynamic mechanical properties of reconstituted networks of purified biopolymers and on understanding how these properties are controlled by factors such as the concentration, stiffness, and structural arrangement of the constituent filaments. We have a comparatively poor understanding of the mechanics of biopolymer-based \textit{composites} constructed with additional material components,  despite their relative ubiquity in the real world. In biological contexts, biopolymer networks generally occupy only a small fraction of the available space, the rest of which might contain a continuous fluid phase as well as inclusions of various types: examples include organelles embedded in the cellular cytoskeleton \cite{fletcher_cell_2010,blanchoin_actin_2014,bonucci_how_2023}, red cells and platelets in the fibrin network of a blood clot \cite{weisel_enigmas_2008,lam_mechanics_2011, qiu_biophysics_2019}, bacteria in biofilm matrices \cite{flemming_biofilm_2010, nadell_extracellular_2015, flemming_biofilm_2023}, and a wide range of cell types and cellular aggregates in the collagenous extracellular matrix \cite{cox_matrix_2021}. Tissues, too, are generally well described as soft, biopolymer-based composites. In order to understand the mechanobiology of these materials and learn how to design composite biomaterials with desired properties that mimic or improve upon those observed in living things \cite{guimaraes_stiffness_2020}, we require an understanding of the mechanical interplay between biopolymer networks and their inclusions. 
 
Unlike typical elastic materials, biopolymer networks tend to deform in a manner that is highly nonaffine, or locally heterogeneous, such that the macroscopic and microscopic deformation fields differ. Consequently, the viscoelastic response of a network can sensitively depend on the length scale over which a deformation is applied \cite{head_distinct_2003,head_mechanical_2005,heussinger_floppy_2006,shahsavari_size_2013,broedersz_modeling_2014, yang_local_2023}. In other words, the apparent stiffness in such networks is highly length-scale dependent. By controlling the characteristic length scale over which deformations occur within such a material, one can in principle produce dramatic changes in the macroscopic mechanical response. As we will explore in this work, in certain varieties of composite materials, this characteristic deformation length scale is a well-defined function of the composition. Specifically, we will consider composites with rigid spherical inclusions, in which, under a macroscopically applied strain, the typical length scale over which local strains are applied to the interstitial material can be identified as the typical distance between neighboring inclusions \cite{davis_elastic_1994}, a function of the particle radius and volume fraction. As the volume fraction increases, the typical inter-inclusion distance decreases. Thus, to describe the dependence of the overall composite stiffness on the volume fraction of inclusions, we must account for both the conventional ``Einstein" stiffening (linear in the volume fraction in the dilute limit) \textit{and} the dependence of the interstitial matrix's elasticity on the typical strain length scale, which itself is a function of the inclusion volume fraction. In such a system, the degree to which the elasticity of the interstitial matrix depends on the typical deformation length scale matters greatly, and can in principle give rise to exotic behaviors in the composite, such as extreme tunability of the elastic modulus at unusually low volume fractions.

\section{Scaling theory} \label{sec:scalingtheory}

\subsection{Volume fraction-dependent stiffening in ordinary composites} \label{sec:ordinary}

Before considering the case of a disordered network, we first imagine that the interstitial material is a simple linear elastic material, with a shear modulus $G_0$ that is independent of both the level of applied strain and the scale on which the strain is applied. For a simple composite consisting of an ordinary elastic material with rigid spherical inclusions of uniform size, we can write the dependence of the effective or apparent shear modulus $G(\phi)$ of the composite as
\begin{equation}
    G = A(\phi) G_0,
\end{equation}
in which $G_0$ is the linear shear modulus of the interstitial medium and $A(\phi)$ is a volume fraction-dependent ``stiffening factor'' that satisfies $A(0)=1$. Crucially, in ordinary composites, $G_0$ itself does not depend on $\phi$. 

Various expressions have been proposed for the stiffening factor $A(\phi)$. In the dilute regime, for rigid $d$-spheres embedded in a matrix of shear modulus $G_0$ and Poisson's ratio $\nu_0$, an effective medium approach described in Ref. \cite{torquato_random_2002} yields the following expression:
\begin{equation} \label{eq:stiffeningfactor}
    A(\phi) = \frac{1}{1-\phi}\left(1+\dfrac{d(1+\nu_0)}{4}\phi\right).
\end{equation}
If one enforces incompressibility of the interstitial matrix ($\nu_0 = 1$ for $d=2$ or $\nu_0 = 1/2$ for $d=3$), a first-order Taylor expansion near $\phi=0$ recovers the $d$-dimensional Einstein result, $A(\phi) = 1 + (1 + d/2)\phi$ \cite{brady_einstein_1984}. In disordered networks with bending interactions, $\nu_0$ has been shown to depend on the network architecture, connectivity $z$, and bending rigidity $\kappa$ \cite{shivers_nonlinear_2020}. 
For packing-derived networks with low $\kappa$, observed values are typically positive and below the two-dimensional upper bound of $1$. Thus, for simplicity, in analyzing our simulation results, we will assume a constant Poisson's ratio of $\nu_0 = 1/3$, consistent with a two-dimensional Cauchy solid with central force interactions only  \cite{seung_defects_1988}. In the end, the chosen value of $\nu_0$ has a negligible effect on our results.

\subsection{Volume fraction-dependent stiffening in composites with length-scale dependent elasticity}  \label{sec:lengthscaledependent}

Motivated by the observation of length-scale-dependent elasticity in biopolymer networks, we now consider the elasticity of a system for which the apparent shear modulus of the matrix depends on the length scale over which the deformation is applied, which we denote $\lambda_d$. That is, $G_0 = G_0(\lambda_d)$. For an inclusion-free sample of such a material under a macroscopically applied strain, the relevant length scale is the size of the sample, $\lambda_d = L$. In this case, the apparent shear modulus of the sample is given by $G_0(L)$. 

\begin{figure}[htb!]
  \centering
  \includegraphics[width=1\columnwidth]{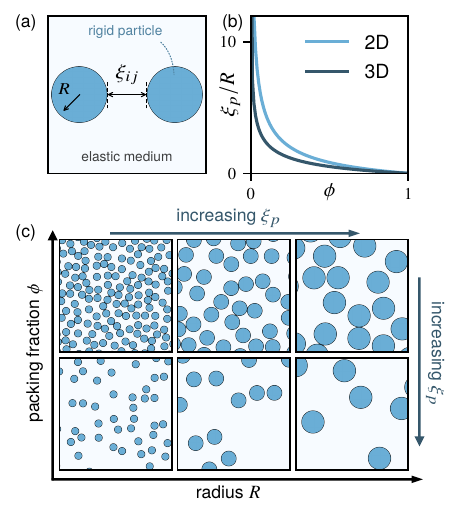}
  \caption{\label{fig:schematic_with_phi_vs_R}
    \textbf{Inclusions restrict the typical size of unobstructed regions of an elastic medium.} We define this size as the average edge-to-edge spacing $\xi_p\equiv \langle \xi_{ij} \rangle$ between pairs of Voronoi neighbors $i$ and $j$, as sketched in (a). (b)  For uniform spherical particles of radius $R$ and volume fraction $\phi$, the typical edge-to-edge spacing is approximately given by $\xi_p \approx 2R(\phi^{-1/d}-1)$. (c) For constant $\phi$, $\xi_p$ increases with increasing $R$. For constant $R$, $\xi_p$ decreases with increasing $\phi$.
  }
\end{figure}

However, for a sample with rigid inclusions, strains imposed on the interstitial network are not determined by the macroscopically applied deformation, but instead are primarily determined by the relative motion of neighboring inclusions \cite{davis_elastic_1994}. In this case, the relevant length scale is instead the typical size of the unobstructed regions of the matrix, which we denote $\xi_p$.  For a network with rigid $d$-dimensional spherical inclusions of uniform size, the typical inter-inclusion spacing $\xi_p$ is approximately given by
\begin{equation} \label{eq:xi_p}
    \xi_p \equiv \langle \xi_{ij} \rangle \approx 2R\left(\phi^{-1/d} - 1\right).
\end{equation}
in which $\xi_{ij}$ is the nearest distance between the edges of particles $i$ and $j$, as shown in Fig. \ref{fig:schematic_with_phi_vs_R}a, and the average is taken over all pairs of Voronoi neighbors.
In Fig. \ref{fig:schematic_with_phi_vs_R}b, we plot $\xi_p/R$ as a function of particle volume fraction $\phi$ for both $d=2$ and $d=3$. In Fig. \ref{fig:delaunay_fig}, we demonstrate that this expression is consistent with the average inter-inclusion spacing determined from the Delaunay triangulation of particle positions in 2D systems. 
Equation \ref{eq:xi_p} indicates that for a given particle volume fraction $\phi$, the typical inter-inclusion spacing $\xi_p$ increases with increasing particle radius $R$. For a fixed particle radius $R$, the inter-inclusion spacing $\xi_p$ decreases with increasing particle volume fraction $\phi$. This behavior is shown schematically in Fig. \ref{fig:schematic_with_phi_vs_R}c. In this case, if the interstitial material has a length-scale dependent elasticity, we expect the apparent shear modulus of the interstitial matrix to depend on the typical inter-inclusion spacing $\xi_p$, i.e. $G_0 = G_0(\xi_p)$.
Incorporating the stiffening factor $A(\phi)$, we would then expect the shear modulus of the composite to behave as
\begin{equation}
    G = A(\phi) G_0(\xi_p).
\end{equation}

\subsection{Scaling argument for volume fraction-dependent stiffening in critical networks due to finite size effects}  \label{sec:scaling}

For disordered networks, length-scale-dependent elasticity becomes especially apparent near mechanical critical points \cite{broedersz_criticality_2011,shivers_scaling_2019,arzash_finite_2020}. Here we will focus on the behavior of networks near the central force rigidity transition, which occurs at the isostatic connectivity $z_c=2d$. Another possibly relevant critical point is the bending rigidity transition, which occurs at significantly lower connectivities. For simplicity, we restrict our focus to networks in the vicinity of the central force rigidity transition.

In the thermodynamic limit (i.e., for large systems), the linear shear modulus of networks with connectivity $z$ above the central force isostatic point $z_c$ scales with $\Delta z = z - z_c$ as $G_0 \propto (\Delta z)^f$,
while on either side of the transition the nonaffinity scales as $\Gamma_0 \propto |\Delta z|^{-\lambda}$ and the correlation length scales as $\xi \propto \xi_{\pm}|\Delta z|^{-\nu}$.
In a finite-sized system, the divergence of the correlation length is suppressed by the finite linear size $L$ of the network,
\be
\xi_\mathrm{max} = \xi_{\pm}|\Delta z|^{-\nu}\sim L \quad \text{for} \quad \phi=0.
\ee
Thus for inclusion-free networks ($\phi = 0$), we expect the linear modulus $G_0$ to exhibit a system-size dependence captured by \cite{broedersz_criticality_2011}
\be
G_0 \equiv G \big|_{\phi=0}= L^{-f/\nu} \mathcal{F}_\pm \left( \Delta z L^{1/\nu} \right).
\ee
For the nonaffinity, one finds 
\be
\Gamma_0 \equiv \Gamma\big|_{\phi =0} = L^{\lambda/\nu} \mathcal{G}_\pm \left(\Delta zL^{1/\nu}\right).
\ee

For systems with inclusions, we expect the typical inter-inclusion spacing $\xi_p$ to act as an upper limit on the correlation length of the interstitial matrix,
\be
\xi_\mathrm{max} = \xi_{\pm}|\Delta z|^{-\nu}\sim \xi_p \quad \text{for} \quad \phi>0,
\ee
since $\xi_p < L$.
Because $\xi_p$ acts as an upper bound on the correlation length just as the system size $L$ does in the absence of inclusions, we expect $G = A(\phi) G_0(\xi_p)$ in the presence of inclusions. Thus, we expect the linear shear modulus to exhibit a scaling behavior of the form
\be \label{eq:G_scaling}
G =  A(\phi)\xi_p^{-f/\nu}\mathcal{F}_\pm \left(\Delta z \xi_p^{1/\nu}\right)
\ee
in which $f$ and $\nu$ are the same scaling exponents as in the inclusion-free case, and $\mathcal{F}_\pm$ is the same scaling function as in the inclusion-free case. Near the critical connectivity $z_c$, we have $\mathcal{F}_\pm \sim \text{const}$. For small $\phi$, the inter-particle spacing decreases with volume fraction as $\xi_p \propto \phi^{-1/d}$, so that $G\sim A(\phi) \phi^{f/d\nu}$.  Consequently, near the critical point $z_c$ and for small $\phi$, the excess stiffness is predicted to scale as 
\begin{equation}
    G - G_0 \propto \phi^\eta \quad \text{with} \quad \eta = 1 + \frac{f}{d\nu}.
\end{equation}
for small $\phi$. This power-law scaling differs from the standard Einstein relationship, in which the excess stiffness scales as $G - G_0 \propto \phi$ in the dilute regime. However, we should note here that our argument for the interparticle distance $\xi_p$ acting as a finite size limit on network correlations presumably fails in the extreme dilute (low-$\phi$) limit, in which infinitesimally small and distant particles may act as small, far-field defects in the matrix. In this case we would only expect the predicted low-$\phi$ scaling relationship to work over an intermediate range of $\phi$ values, above the extreme dilute limit. This is indeed what we observe, as is shown in Fig. \ref{fig:G_scaling_dilute}.

For the nonaffinity, we hypothesize a form
\begin{equation}
    \Gamma = B(\phi) \Gamma_0
\end{equation}
in which $B(\phi)$ is a \textit{nonaffinity amplification factor} that satisfies $B(\phi = 0) = 1$. Note that $B(\phi)$ is some function of the volume fraction that describes the additional nonaffinity due to the presence of inclusions, even for an interstitial mesh deforming as homogeneously as possible given the constraints of the inclusions. We leave the determination of this function for future work. Incorporating the $\xi_p$-limited scaling of $\Gamma_0$, we find
\be \label{eq:Gamma_scaling}
\Gamma = B(\phi) \xi_p^{\lambda/\nu} \mathcal{G}_\pm \left(\Delta z \xi_p^{1/\nu}\right)
\ee
in which $\lambda$ is the same nonaffinity scaling exponent as in the inclusion-free case. The nonaffinity is expected to exhibit a maximum at the $\xi_{p}$-dependent critical connectivity, $z_c^*$ (possibly just include this point later).

\section{Simulations} \label{sec:simulations}

To test the proposed relationships, we perform simulations of two-dimensional disordered networks with embedded rigid inclusions. In this section, we describe our simulation protocol.

First, we prepare periodic packing-derived networks with connectivity $z$ following standard protocols described in prior work \cite{wyart_elasticity_2008}. The initial network structure is derived from the contact network of a densely jammed packing of soft, radially bidisperse disks; that is, the centers of disks correspond to network nodes and the contacts between disks correspond to network edges. The initial connectivity is $z_0 \approx 5.5$. The connectivity is then reduced to the desired value $z$ via a biased dilution procedure. To minimize spatial variation in $z$, we follow the biased dilution protocols described in Refs. \cite{wyart_elasticity_2008,rens_rigidity_2019}. This is important for our purposes to ensure that the network regions between particles can be properly thought of as portions of a network with connectivity $z$.  Network edges, which will correspond to springs in the mechanical network, are assigned a linear 1D modulus $\mu$ and rest length $\ell_{0,ij}$. The stretching energy of the network is given by
\begin{equation}
    E_\mathrm{s} = \frac{1}{2} \mu \sum_{\langle ij \rangle} \frac{1}{\ell_{ij,0}} \left( \ell_{ij} - \ell_{0,ij} \right)^2,
\end{equation}
in which the sum is taken over all pairs of connected nodes $i$ and $j$.  Pairs of adjacent springs are assigned a harmonic bending interaction with stiffness $\kappa$. The total bending energy of the network is given by
\begin{equation}
    E_\mathrm{b} = \frac{1}{2} \kappa \sum_{\langle ijk \rangle} \frac{1}{\ell_{ijk,0}}\left( \theta_{ijk} - \theta_{0,ijk} \right)^2,
\end{equation}
in which the sum is taken over all triplets of connected nodes $i$, $j$, and $k$, $\theta_{ijk}$ is the angle between the two edges $\langle ij \rangle$ and $\langle jk \rangle$, and $\ell_{ijk,0} = (\ell_{ij,0} + \ell_{jk,0})/2$. The total energy of the network is then given by
\begin{equation}
    E_\mathrm{tot} = E_\mathrm{s} + E_\mathrm{b}.
\end{equation}

After the network is prepared as described above, disk-shaped inclusions of radius $R$ are randomly placed throughout the network at positions $\left\{\mathbf{x}_i\right\}$ with a chosen volume fraction $\phi$. Note that, as we will consider only volume-preserving simple shear in this work, $\phi$ remains constant throughout each mechanical testing step (it is not a function of strain).  When an inclusion is placed, any spring that intersects with the surface of the placed inclusion is connected to the surface at the point of intersection with a freely-rotating pin joint, and all nodes and springs inside the radius of the newly placed inclusion are discarded. Inclusions are treated as rigid objects and are thus free to translate and rotate, subject to the forces exerted by the attached springs. To achieve this, each inclusion $i$ is assigned an orientation vector $\hat{\mathbf{n}}_i = (\cos\vartheta_i, \sin\vartheta_i)$, in which $\vartheta_i$ is the angle of orientation relative to the $x$-axis. We define $\vartheta_i = 0$ in the initial, unstrained configuration. The relative angular positions of the surface nodes with respect to the node's orientation vector are fixed according to the initial configuration, i.e. each node $j$ attached to the surface of inclusion $i$ is assigned an angular position $\vartheta_{ij} = \vartheta_i + \Delta\vartheta_{0,ij}$, in which $\Delta\vartheta_{0,ij}$ is the angular position of node $j$ with respect to the orientation vector in the initial configuration. Because the inclusions are rigid objects, the values of $\{\Delta\vartheta_{0,ij}\}$ remain fixed. Then, given the position $\mathbf{x}_i$ and orientation vector $\hat{\mathbf{n}}_i$ of inclusion $i$, the positions of all of the surface nodes are uniquely determined. It is important to note that we make a distinction between network nodes, which have freely variable positions, and inclusion surface nodes, which have positions that are determined by the position and orientation of the inclusion. The degrees of freedom of the system are then the positions of the network nodes, the positions of the inclusions, the orientations of the inclusions, and the components of the macroscopic strain tensor $\bm{\Lambda}$. For simple shear strain $\gamma$ in 2D,
\begin{equation}
    \bm{\Lambda} = \begin{pmatrix}
    1 & \gamma \\
    0 & 1
    \end{pmatrix}.
\end{equation}
Because inclusion overlaps do not occur in the small strain regime for the range of $\phi$ values considered here, we do not consider steric interactions between inclusions. A portion of a simulated network with embedded inclusions is shown in Fig. \ref{fig:3D_schem}a.

Macroscopic strain is applied using Lees-Edwards boundary conditions \cite{lees_computer_1972}. 
The rest lengths and rest angles are defined such that the initial configuration of the network has zero energy. To measure the shear modulus $G$ under conditions of quasistatic deformation, we impose a small shear strain $\gamma = 0.0001$ and numerically minimize the total energy, 
\begin{equation}
   E(\gamma) = \underset{\{\mathbf{x}_i\}_\mathrm{net}, \{\mathbf{x}_j, \vartheta_j\}_\mathrm{inc}}{\mathrm{min}} \; E_\mathrm{tot}(\underbrace{\{\mathbf{x}_i\}_\mathrm{net}}_{\substack{\text{network} \\ \text{nodes}}},\underbrace{\{ \mathbf{x}_j, \vartheta_j\}_\mathrm{inc}}_{\text{inclusions}}, \gamma)
\end{equation}
subject to the fixed applied strain $\gamma$. We perform this minimization using the \texttt{L-BFGS} algorithm \cite{nocedal_numerical_2006}. Above, $\{\mathbf{x}_i\}_\mathrm{net}$ refers to the list of positions of the network nodes, and $\{\mathbf{x}_j, \vartheta_j\}_\mathrm{inc}$ refers to the list of positions and orientations of the inclusions. After minimization, we compute the linear shear modulus as
\begin{equation}
    G = 2 \Delta E / \gamma^2,
\end{equation}
in which $\Delta E = E(\gamma) - E(0) = E(\gamma)$. In practice, we apply both positive and negative strain to avoid artifacts caused by asymmetry in the finite-sized random network structures. Equivalently, we consider an ensemble of configurations including each network and its left-right mirror image. To mitigate effects related to finite system size, we perform simulations with $L = 120 \ell_0$ and separately verify that our results correspond to a regime of sufficiently large size for system-size related effects to be negligible.

We compute the nonaffinity $\Gamma$ as
\begin{equation}
    \Gamma = \lim_{\gamma \to 0} \frac{1}{\ell_0^2 \gamma^2} \bigg\langle \left| \mathbf{u}_i^\mathrm{NA} (\gamma)\right|^2 \bigg\rangle,
\end{equation}
in which $\langle \cdot \rangle$ indicates an average over all network nodes and $\mathbf{u}_i^\mathrm{NA} = \mathbf{u}_i - \mathbf{u}_i^\mathrm{aff}$ is the nonaffine displacement of node $i$. Here, $\mathbf{u}_i = \mathbf{x}_i - \mathbf{x}_{i,0}$ is the actual displacement of node $i$ between the initial and (relaxed) strained configurations, and $\mathbf{u}_i^\mathrm{aff}  = \mathbf{\Lambda} \mathbf{x}_{i,0}$ corresponds to the displacement of node $i$ under an affine applied strain $\gamma$. Intuitively, the nonaffinity measures how nonuniformly the network deforms under the applied strain. Note that another potentially interesting measure of nonaffinity is that of the displacements of the inclusions themselves. We have not considered this quantity here.

\section{Results and discussion} \label{sec:results}

To test the proposed relationships, we first measure the relevant scaling exponents in simulations of disordered networks in the absence of inclusions (see Appendix A), for which we determined $f=1.0$, $\varphi = 2.0$, and $\nu = 1.0$ for the variety of network structure chosen here. We then perform simulations of disordered networks with rigid inclusions.

\begin{figure}[htb!]
    \centering
    \includegraphics[width=1\columnwidth]{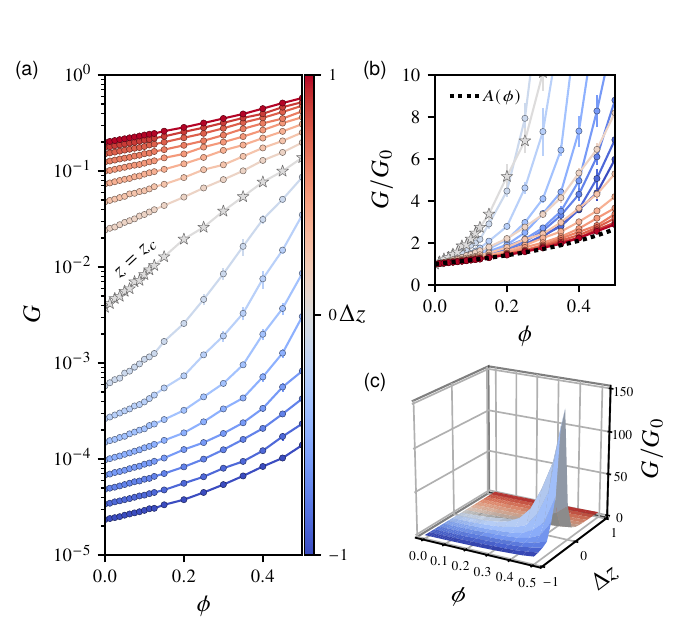}
    \caption{\label{fig:G_vs_phi_with_varying_z}
    \textbf{Criticality strongly enhances the reinforcement of disordered networks by rigid inclusions. } (a) The linear shear modulus $G$ plotted as a function of inclusion volume fraction $\phi$ for networks with varying connectivity $z$. Close to the critical connectivity $z$, $G$ exhibits an anomalously strong dependence on $\phi$ due to correlation length suppression. (b) Normalized shear modulus $G/G_0$, in which $G_0$ corresponds to the stiffness of the background matrix ($\phi=0$) at each value of $z$. The dotted black line corresponds to the ordinary stiffening factor $A(\phi)$, which captures the stiffening behavior in networks far from the critical point. (c) The same data as in (b), plotted as a surface to emphasize the enhanced stiffening near $z_c$. Here $\kappa = 10^{-5}$ and each point represents an average over $n_\mathrm{s}=5$ samples. Error bars represent $\pm 1$ standard deviation.
    }
\end{figure}

We first test the influence of the interstitial matrix connectivity on the volume-fraction-dependence of the composite shear modulus. In Fig. \ref{fig:G_vs_phi_with_varying_z}a, we plot the measured composite shear modulus $G$ as a function of particle volume fraction $\phi$ for systems in which the network connectivity varies from $z = 3$ to $z = 5$, encompassing the two-dimensional isostatic connectivity $z_c = 4$. This range corresponds to $\Delta z \equiv z - z_c \in [-1, 1]$. We find that, far from the critical strain, the composite shear modulus exhibits a dependence on $\phi$ that is consistent with the ordinary stiffening factor $A(\phi)$ over a considerable range of volume fractions $\phi$. This is especially true above the critical strain. To make this effect more evident, in Fig. \ref{fig:G_vs_phi_with_varying_z}b we plot the normalized shear modulus $G/G_0$ for the same data, along with the result expected for a simple linear medium $G/G_0 = A(\phi)$. Remarkably, networks with connectivity near the critical point exhibit a much stronger dependence of the composite shear modulus on $\phi$ than expected from the ordinary stiffening factor $A(\phi)$. In other words, the stiffening effect of increasing the particle volume fraction is greatly enhanced by the critical length-scale-dependent elasticity of the interstitial network near the isostatic point. This is especially evident in Fig. \ref{fig:G_vs_phi_with_varying_z}c, in which we plot the normalized shear modulus surface as a function of $\phi$ and $\Delta z$ to emphasize the enhanced stiffening near the critical point.

\begin{figure}[htb!]
\centering
\includegraphics[width=1\columnwidth]{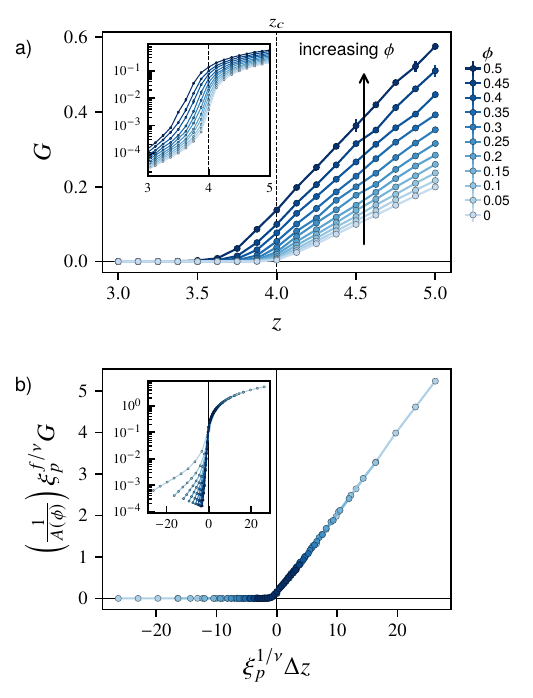}
\caption{\label{fig:G_vs_z_with_varying_phi}
\textbf{ Rigid inclusions increase the linear shear modulus $G$ in part by suppressing the network correlation length in a manner consistent with effective finite size scaling.}  Here $\kappa = 10^{-5}$ and each point represents an average over $n_\mathrm{s}=5$ samples. (a) Linear shear modulus $G$ vs. network connectivity $z$ for systems with varying inclusion volume fraction $\phi$, with inclusion radius $R \approx 3.8 \ell_0$. (b) When rescaled according to the scaling prediction given in Eq. \ref{eq:G_scaling}, the data collapse onto a single curve. Inset: A log-linear plot of the same data. Error bars represent $\pm1$ standard deviation.
}
\end{figure}

We next test the scaling predictions provided in Eqs. \ref{eq:G_scaling} and \ref{eq:Gamma_scaling}. In Fig. \ref{fig:G_vs_z_with_varying_phi}a, we replot the data from Fig. \ref{fig:G_vs_phi_with_varying_z}, now with the measured composite shear modulus $G$ as a function of network connectivity $z$ for systems with varying particle volume fraction $\phi$. We find that increasing the volume fraction of particles leads to a downward shift in the apparent critical connectivity $z_c^*$. As we will discuss later in this section, this behavior is in agreement with the prediction given in Eq. \ref{eq:zc_scaling}. To be more quantitative, we plot the same data as in Fig. \ref{fig:G_vs_z_with_varying_phi}a rescaled according to the scaling prediction given in Eq. \ref{eq:G_scaling}. We find that the data collapse onto a single master curve, consistent with the scaling prediction. In the inset of Fig. \ref{fig:G_vs_z_with_varying_phi}b, we plot the same data on a log-linear scale.

\begin{figure}[htb!]
  \centering
  \includegraphics[width=1\columnwidth]{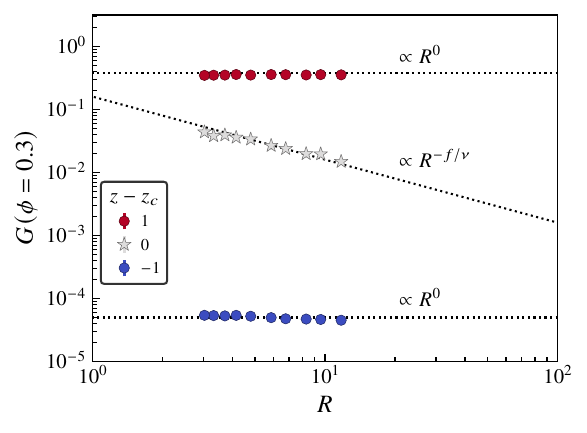}
  \caption{\label{fig:G_vs_R} \textbf{Near criticality, the linear shear modulus $G$ exhibits a power law dependence on particle radius $R$, consistent with the predictions of the scaling theory.} These data correspond to systems with linear system size $L = 120 \ell_0$ and inclusion volume fraction $\phi = 0.3$. Here $\kappa = 10^{-5}$ and each point represents an average over $n_\mathrm{s}=5$ samples. Error bars representing $\pm1$ standard deviation are smaller than the markers.
 }
 \end{figure}

One remarkable consequence of the proposed dependence of the composite elasticity on $\xi_p$ is the prediction that the composite shear modulus should exhibit a dependence on the particle radius $R$ near the critical connectivity, as $G \propto R^{-f/\nu}$. This is unusual, as in typical composites it is only the volume fraction, and not the particle size, that influences the overall stiffness. In Fig. \ref{fig:G_vs_R}, we plot the measured composite shear modulus $G$ as a function of particle radius $R$ for systems with a fixed particle volume fraction $\phi = 0.3$. We find that the composite shear modulus exhibits a power law dependence on the particle radius $R$ with a decay of $G \propto R^{-f/\nu}$, consistent with the predictions of the scaling theory, using the exponents $f$ and $\nu$ determined from inclusion-free simulations. For values of $z$ far from the critical point, we observe that $G$ is independent of $R$, consistent with the expected behavior for a simple linear elastic medium.

\begin{figure}[htb!]
\centering
\includegraphics[width=1\columnwidth]{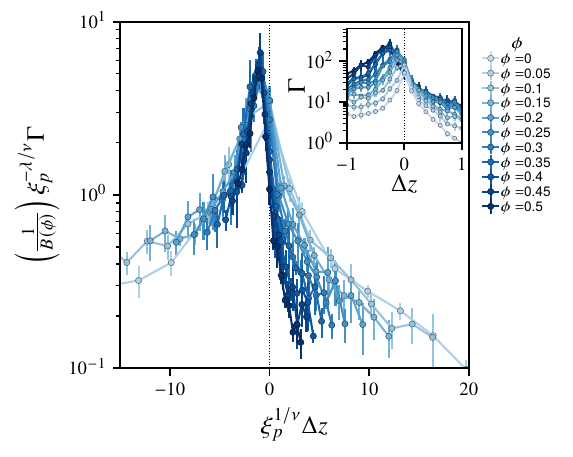}
\caption{\label{fig:nonaffinity}
\textbf{Increasing the particle volume fraction shifts the peak in the network nonaffinity in a manner consistent with effective finite size scaling.} In agreement with the scaling theory, the peak in the nonaffinity $\Gamma$ shifts to lower values of network connectivity $z$ as the particle volume fraction $\phi$ is increased. Inset: Measurements of nonaffinity $\Gamma$ as a function of distance to the critical connectivity $\Delta z = z - z_c$ for systems with varying particle volume fraction $\phi$. Main panel: We observe reasonable collapse upon rescaling $\Gamma$ and $\Delta z$ according to the scaling ansatz (Eq. \ref{eq:Gamma_scaling}) if we choose, for the empirical nonaffinity amplification function,  $B(\phi) = 1 + c \phi^2$, with $c = 50$. Error bars represent $\pm 1$ standard deviation.
}
\end{figure}

In the inset of Fig. \ref{fig:nonaffinity}, we plot the nonaffinity $\Gamma$ as a function of network connectivity $z$ for systems with varying inclusion volume fraction $\phi$. For a given value of the particle volume fraction $\phi$, the nonaffinity $\Gamma$ reaches a maximum at the corresponding apparent critical connectivity $z_c^*$, the value of which shifts to lower values of $z$ as the particle volume fraction is increased. This is qualitatively consistent with the finite size scaling-like behavior predicted by the scaling theory, i.e. that decreasing the typical inter-particle spacing should shift the apparent critical connectivity. For inclusion-free networks, the critical connectivity shifts with finite system size $L$ as \cite{broedersz_criticality_2011} $z_c^* = z_{c,\infty}-a L^{-1/\nu}$, in which $z_{c,\infty} \equiv z_c(L\to\infty)$. For networks with inclusions, the correlation length is limited by the typical spacing between inclusion edges, $\xi_p$, so we expect the critical connectivity to depend not on the system size but rather on the typical inter-inclusion spacing, as
\be \label{eq:zc_scaling}
z_c^* = z_{c,\infty} - b \xi_p ^{-1/\nu}
\ee
in which $b$ is a fit parameter and $\nu$ is the same correlation length exponent applicable to the inclusion-free case. This shift is expected to be negative, yielding a reduction in $z_c$, because for inclusion-free networks, decreasing the system size $L$ results in a decrease in the apparent critical connectivity. In the main panel of Fig. \ref{fig:nonaffinity}, we rescale $\Delta z$ by $\xi_p^{-1/\nu}$ and find that the locations of the peaks reasonably coincide. To properly test the scaling prediction in Eq. \ref{eq:Gamma_scaling}, we need to know the nonaffinity amplification factor $B(\phi)$. Unfortunately, we do not know this function. Empirically, we find that the function $B(\phi) = 1 + c \phi^2$, in which $c$ is a fit parameter, produces a reasonable collapse near the critical point for nonzero $\phi$, as shown in Fig. \ref{fig:nonaffinity}.

\begin{figure}[htb!]
    \centering
    \includegraphics[width=1\columnwidth]{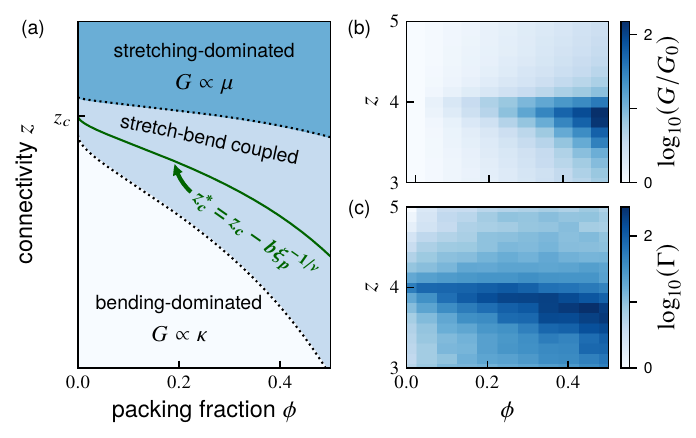}
    \caption{\label{fig:phase_diagrams}
    \textbf{Phase diagrams for the shear modulus and nonaffinity reveal a shifted critical strain, enhanced reinforcement, and large nonaffine rearrangement.} (a) Schematic phase diagram for the composite stiffness as a function of connectivity $z$ and particle volume fraction $\phi$, with the shifted phase boundary between the bending-dominated and stretching-dominated regimes indicated by a thick line.  The boundary shifts due to finite size effects imposed by the inter-particle spacing $\xi_p$, resulting in $z_c^* = z_{c,\infty} - b \xi_p^{-1/\nu}$ where $z_{c,\infty} = 4$. (b) The normalized shear modulus $G/G_0$ increases near the boundary as the particle volume fraction $\phi$ increases. (c) The nonaffinity $\Gamma$ grows as the boundary is approached from below and above.
    }
    
\end{figure}

As noted above, increasing the volume fraction leads to a downward shift of the critical connectivity $z_c^*$ as the particle volume fraction $\phi$ is increased. In Fig. \ref{fig:phase_diagrams}a, we sketch the corresponding phase diagram as a function of particle volume fraction $\phi$ and network connectivity $z$. The phase boundary  separating the bending-dominated ($G \propto \kappa$) and stretching-dominated ($G \propto \mu$) mechanical regimes shifts to lower values of connectivity $z$ as the particle volume fraction $\phi$ increases. In Fig. \ref{fig:phase_diagrams}b, we plot the relative stiffness $G/G_0$ as a function of connectivity $z$ and volume fraction $\phi$. Here, we see that the effect of increasing $\phi$ on the relative stiffness of the composite increases dramatically in the regime around the phase boundary. This is consistent with the predictions of the scaling theory and suggests that critical behavior in the interstitial network can dramatically enhance the reinforcement effect of rigid inclusions on disordered network elasticity. The notion that the phase boundary is shifted is also supported by the observed shift in the peak in $\Gamma$ as $\phi$ is increased, which is clearly evident in Fig. \ref{fig:phase_diagrams}c.

The observed trends will of course fail at larger values of the particle volume fraction, i.e. in the regime near jamming ($\phi_j \approx 0.84$ in 2D). Above the jamming threshold, the composite's mechanics will become controlled by the rigidity of the inclusions. For values of $\phi$ below but close to the jamming regime, one would expect to see critical behavior akin to that described in Ref.  \cite{zhao_elasticity-controlled_2024}.

\section{Conclusion}

We have shown that the presence of rigid inclusions in disordered networks can give rise to behavior akin to finite size effects in inclusion-free networks: the elastic correlation length of the network is suppressed by the typical inter-inclusion length scale, beyond which correlations are effectively screened. This characteristic length scale is a function of the particle size and volume fraction. Using a scaling ansatz that accounts for this behavior, we are able to make quantitative predictions for the volume fraction-dependent stiffening of networks with embedded inclusions in the vicinity of the phase transition; in particular, we predict and demonstrate by simulations an unusual dependence of the composite elasticity on both the size and volume fraction of inclusions present, in striking contrast to the long-established result for filled composites with conventional linear elastic matrices, for which only the volume fraction plays a role in setting the composite's elasticity. For disordered network composites in the dilute regime of low inclusion volume fraction $\phi$, our theory predicts a scaling of the excess shear modulus of $G(\phi) - G_0 \propto \phi^{\eta}$ with $\eta = 1 + f/(d\nu)$, in which  $d$ is the dimensionality and $f$ and $\nu$ are, respectively,  the exponents describing the scaling of the network stiffness $G_0$ and correlation length $\xi$ near the isostatic critical point. Our simulations show that this prediction is satisfied over an intermediate range of $\phi$ values but fails in the extreme dilute (low-$\phi$) regime, where the argument for treating the inter-inclusion distance as an effective network size becomes invalid. Far from the critical point, where the network correlation length is smaller than the typical distance between inclusions, we observe the typical Einstein dependence of the stiffness on volume fraction, analogous to the dependence of the viscosity of suspensions on particle density \cite{einstein_neue_1906,einstein_berichtigung_1911}.  

Our results suggest that criticality could provide enhanced tunabiility of the mechanics of composite materials, suggesting both interesting biological implications and possible strategies for designing functional engineered tissues \cite{gosselin_designing_2018} and highly tunable synthetic composites. These ideas may also be useful in other domains in which the mechanics of network-based composites are important, e.g. in food science \cite{vilgis_soft_2015, zielbauer_networks_2017,macqueen_muscle_2019, mathijssen_culinary_2023}.  Importantly, our results emphasize the need to consider the interplay between critical behavior and geometric constraints in predicting the behavior of disordered network-based composites.  While we have only considered two-dimensional systems in this work, the same scaling relationships derived here should in principle apply to three dimensional systems and should be readily testable using existing simulation methods. 

One interesting consequence of the behavior we have discussed here is that signatures of criticality are apparent at much lower connectivity when inclusions are present. Related behavior has been shown to occur in networks with angle-constraining crosslinks \cite{das_redundancy_2012}. Biological networks are evidently far below the three-dimensional isostatic point (with average connectivity $z \approx 3 \ll z_c$ in which $z_c = 6$), which would lead one to believe that the relevance of criticality to their linear elasticity is minimal \footnote{It is important to note, however, that the same is not true for strain-controlled criticality.}. However, in biological contexts, networks are rarely inclusion-free. The presence of cells and other inclusions in biological networks would, we expect, tend to reduce the maximum correlation length, such that the apparent isostatic point shifts to lower values of $z$, increasing the effects of criticality in the biologically relevant regime.

Several exciting avenues remain for future work.  For example, investigating systems with internal prestress or prestrain will likely reveal interesting behavior. Tensile prestrain is known to reduce both the critical connectivity and the critical strain required for the onset of stiff, stretching-dominated mechanics in disordered elastic networks \cite{sheinman_nonlinear_2012,arzash_stress-stabilized_2019, hatami-marbini_nonlinear_2021}, while compressive prestrain does the opposite \cite{sheinman_nonlinear_2012, vahabi_elasticity_2016}. Internally generated active tensile prestress, like that generated by molecular motors \cite{koenderink_active_2009, broedersz_molecular_2011,sheinman_actively_2012} or contractile cells \cite{vader_strain-induced_2009,shah_strain_1997,lam_mechanics_2011,pathare_fibrin_2021, qiu_biophysics_2019,zakharov_clots_2024}, can have similar effects . One could also consider, with the tools used here, the case in which the inclusions themselves are capable of actively exerting forces on the surrounding network. In this case, the inclusions would generate prestress that stiffens (or possibly softens \cite{sarkar_unexpected_2024}) the interstitial network  \cite{goren_elastic_2020, hatami-marbini_mechanical_2021, heyden_strain_2023}, while also reinforcing the network via conventional means.  Notably, these effects can be achieved with forces generated by synthetic inclusions that respond to external signals, e.g., thermoresponsive or magnetoresponsive particles \cite{chaudhary_exploiting_2020, chaudhary_thermoresponsive_2019, sarkar_unexpected_2024}. 

It would be interesting to investigate the impact of relaxing certain assumptions made in this work. For example, one could consider how different spatial arrangements of particles \cite{torquato_random_2002}, e.g. fractal aggregates \cite{straube_determination_1992,huber_universal_1999,huber_mechanism_2002,kluppel_role_2003} and different distributions of particle size and shape might influence the suppression of the network correlation length in a manner different from that described here. Alternatively, one could consider systems with different interactions between the particles and the matrix beyond permanent binding, such as attractive interactions \cite{dellatolas_local_2023}, which will influence the volume fraction dependence of the overall composite stiffness. Exploring other varieties of underlying network structure, such as lattice-based and Mikado networks, in addition to networks with built-in structural anisotropy, could also produce useful insight. While we have focused on permanent networks in this work, it would be interesting to explore the relevance of correlation length suppression in phenomena involving local network damage or fatigue \cite{steck_multiscale_2023}, which can become especially apparent under the application of large strains, producing a strain-history dependent stiffness due to the  Mullins effect \cite{mullins_thixotropic_1950-1,mullins_stress_1965,song_strain_2023}. In addition, it will be important to understand how correlation length suppression is modified in the case of nonrigid inclusions \cite{islam_random_2019,islam_random_2019-1}, such as liquid droplets \cite{style_stiffening_2015,wei_modeling_2020,rosowski_elastic_2020,lee_mechanobiology_2022,ronceray_liquid_2022,liu_liquidliquid_2023,kothari_crucial_2023,fernandez-rico_elastic_2024,qiang_nonlocal_2024} or otherwise highly deformable particles \cite{gersh_fibrin_2009,van_oosten_emergence_2019}, which are more relevant to real cells and tissues. Furthermore, while our theoretical predictions are limited to systems in which the particle diameter is much larger than the network mesh size, it is not clear how the network correlation length might be influenced if the particle size is comparable to or smaller than the network mesh size \cite{jiang_filled_2023}. This warrants further exploration.

Expanding our work to consider composites in which the elasticity of the interstitial material reflects other mechanical phase transitions, such as bending rigidity percolation ($z \to z_b$) \cite{broedersz_criticality_2011}, strain-driven rigidity percolation ($\gamma \to \gamma_c$) \cite{sharma_strain-controlled_2016, chen_field_2024}, and jamming ($\phi \to \phi_j$) \cite{goodrich_finite-size_2012}, could provide valuable insight. 
While we have only focused on networks near the network elasticity transition corresponding to central force rigidity percolation, the bending rigidity transition is also potentially of interest \cite{broedersz_criticality_2011}. For networks of the sort considered here, the bending rigidity transition coincides with the connectivity percolation transition. Near this transition, the network correlation length would also be expected to diverge in the inclusion-free case, implying that one should also observe suppression of these correlations by $\xi_p$ and enhanced stiffening with power law scaling of $G$ with $\phi$, as we observed here in the central force case.  Another natural next step would be to consider the strain-controlled rigidity transition \cite{sharma_strain-controlled_2016}, near which interesting dynamic critical behavior is also observed \cite{shivers_strain-controlled_2023}. Strain amplification due to the presence of inclusions \cite{mullins_stress_1965, song_guide_2016, song_strain_2023}, when combined with the increased effective connectivity due to the constraints imposed by the inclusions, should in principle lead to a significant reduction in the macroscopic critical strain.  Additionally, examining systems with larger inclusion volume fractions, near contact percolation \cite{shen_contact_2012} or in the densely filled regime where critical behavior controlled by proximity to the jamming transition has been observed \cite{zhao_elasticity-controlled_2024}, could provide additional insight.

Future work should also consider dynamics, which we have ignored here. When a network without inclusions is embedded in a solvent, the divergence of the correlation length and associated growing nonaffine rearrangements near the connectivity-controlled and strain-controlled critical points gives rise to critical slowing down in the stress relaxation \cite{yucht_dynamical_2013,shivers_strain-controlled_2023}, even as the solvent viscosity remains fixed. The presence of inclusions, however, would increase the apparent viscosity of the solvent due to hydrodynamic effects as originally described by Einstein \cite{einstein_neue_1906, einstein_berichtigung_1911}, presumably giving rise to even more dramatic slowing down of stress relaxation. Because of the effect of the inclusions on the critical connectivity and, we assume, the critical strain, this critical slowing down would also occur at lower values of the connectivity than in the inclusion-free case. To properly make predictions about the dynamics, it will be necessary to understand the influence of inclusions on the nonaffinity (i.e., the form of $B(\phi)$), which remains to be determined.

\section{Acknowledgments}

This work was supported in part by the National Science Foundation Division of Materials Research (Grant No. DMR-2224030) and the National Science Foundation Center for Theoretical Biological Physics (Grant No. PHY-2019745). J.L.S. is grateful for the support of the Eric and Wendy Schmidt AI in Science Postdoctoral Fellowship at the University of Chicago.  We would also like to thank the Isaac Newton Institute for Mathematical Sciences, Cambridge, for support and hospitality during the program ``New Statistical Physics in Living Matter,'' where work on this paper was undertaken. This work was supported by EPSRC Grant No. EP/R014604/1.

\appendix

% \clearpage

% \begin{appendices}

\section{Empirical determination of interparticle spacing}

\begin{figure}[htb!]
\centering
\includegraphics[width=1\columnwidth]{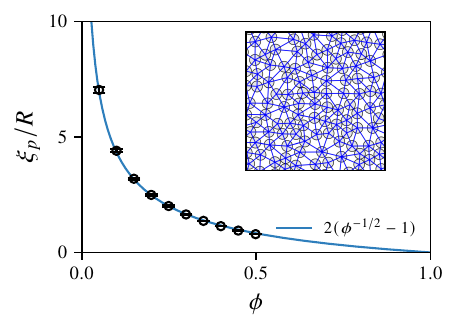}
\caption{\label{fig:delaunay_fig} From the Delaunay triangulation of particle positions, we measure the average inter-inclusion spacing $\xi_p = \langle \xi_{ij} \rangle = \langle r_{ij} - 2R \rangle $, in which $r_{ij}$ is the distance between the centers of inclusions $i$ and $j$ and the averages are taken over all pairs of neighbors in the triangulation. The measured values match the prediction of Eq. \ref{eq:xi_p}. These data correspond to $L=120\ell_0$ and $R = 3.78/\ell_0$ and error bars represent $\pm 1$ standard deviation for 10 samples. Inset: Portion of the Delaunay triangulation of inclusion centers. Edges of the triangulation are colored blue. }
\end{figure}

For a natural definition of inclusion neighbors, we turn to the Delaunay triangulation \cite{lee_two_1980}. The Delaunay triangulation is the dual of the Voronoi diagram, such that vertices connected by an edge in the triangulation correspond to Voronoi neighbors.
To test the validity of the expression for the interparticle spacing, Eq. \ref{eq:xi_p}, we directly measure the average distance between identified neighbors in randomly generated systems. An example of a generated triangulation is shown in the inset of Fig. \ref{fig:delaunay_fig}b. The typical distance between neighboring particle centers is the average length of the edges of the Delaunay triangulation, and the typical distance between particle edges is the same quantity minus twice the radius (see Fig. \ref{fig:schematic_with_phi_vs_R}a). Specifically, we compute the inter-neighbor distance $\xi_{ij} = r_{ij} - 2R$, in which $r_{ij}$ is the center-to-center distance between particles $i$ and $j$ and $R$ is the particle radius. We then compute the average inter-neighbor distance $\xi_p = \langle \xi_{ij} \rangle$ for a given particle configuration. In Fig. \ref{fig:delaunay_fig}b, we plot the average inter-neighbor distance $\xi_p$ as a function of particle volume fraction $\phi$ for systems with $L=120\ell_0$ and $R = 3.78\ell_0$. We find that the measured inter-neighbor distance agrees well with the prediction of Eq. \ref{eq:xi_p}.

\section{Determination of scaling exponents}

\begin{figure}[htb!]
\centering
\includegraphics[width=1\columnwidth]{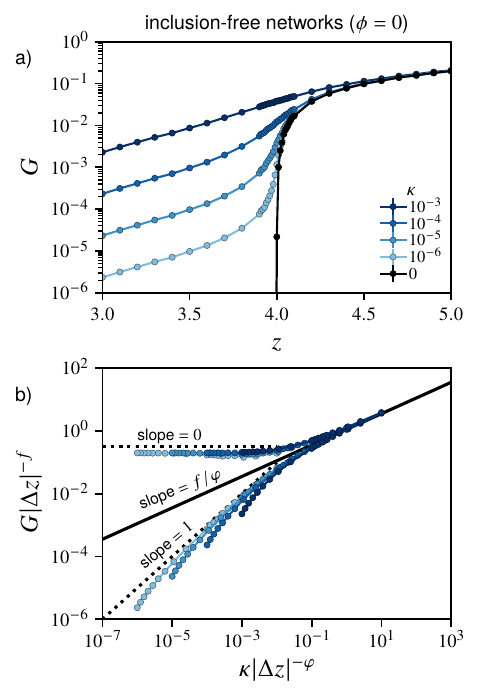}
\caption{\label{fig:empty_scaling}
\textbf{Determination of critical exponents $\bm{f}$ and $\bm{\varphi}$ from inclusion-free networks}. (a) Shear modulus vs. connectivity for 2D packing-derived networks with varying $\kappa$, without inclusions (volume fraction $\phi = 0$). (b) Collapse according to the scaling predictions described in Ref. \cite{broedersz_criticality_2011} with scaling exponents $f=1.0$ and $\varphi = 2.0$. Here $n_\mathrm{samples}=5$ and error bars represent $\pm 1$ standard deviation. }
\end{figure}

\begin{figure}[htb!]
\centering
\includegraphics[width=1\columnwidth]{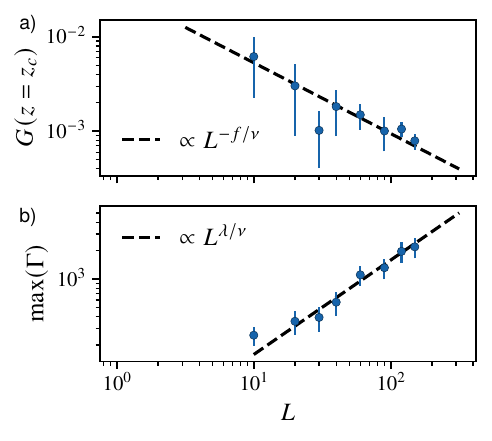}
\caption{\label{fig:empty_fs}
\textbf{Determination of correlation length exponent $\bm{\nu}$ from finite size scaling of inclusion-free networks}. (a) Shear modulus $G$ at the critical connectivity $z_c=4$ for networks without inclusions (volume fraction $\phi = 0$) as a function of linear system size $L$. The dashed line indicates the scaling $G(z=z_c)\propto L^{-f/\nu}$. (b) Maximum nonaffinity $\Gamma$ as a function of linear system size $L$. The dashed line indicates the scaling $\max \Gamma \propto L^{-\lambda/\nu}$. Here we use $f = 1.0$, $\nu = 1.0$, and $\lambda = \varphi - f$, with $\varphi = 2.0$. Here $n_\mathrm{samples}=8$ and error bars represent $\pm 1$ standard deviation. }
\end{figure}

We determine $f$ and $\varphi$ for networks without inclusions (volume fraction $\phi=0$). We expect the $\kappa$- and $z$-dependence of the shear modulus $G_0$ to obey Widom scaling of the form \cite{broedersz_criticality_2011}
\begin{equation}
G_0 = |\Delta z|^f \mathcal{H}_\pm\left(\kappa |\Delta z|^{-\varphi}\right),
\end{equation}
such that a plot of $G_0 | \Delta z|^{-f}$ vs. $\kappa |\Delta z|^{-\varphi}$ should collapse onto a scaling function with a critical branch exhibiting a slope of $f/\phi$. We find that $f = 1.0$ and $\varphi = 2.0$ lead to a reasonable collapse according to this scaling form, as shown in Fig. \ref{fig:empty_scaling}. These values are consistent with prior estimates for similar network structures \cite{wyart_elasticity_2008,rens_micromechanical_2018} and with predictions from effective medium theory \cite{broedersz_criticality_2011,das_redundancy_2012}. We then use $\lambda = \varphi - f$ \cite{broedersz_criticality_2011}. It is important to note that the scaling exponents $f$ and $\varphi$ are not universal and can depend on the network structure \cite{broedersz_criticality_2011}. We estimate the correlation length exponent $\nu=1.0$ from finite size scaling analysis of inclusion-free ($\phi=0$) networks (see Fig. \ref{fig:empty_fs}a-b).

\section{Behavior in the central force limit}

\begin{figure}[ht]
    \centering
    \includegraphics[width=1\columnwidth]{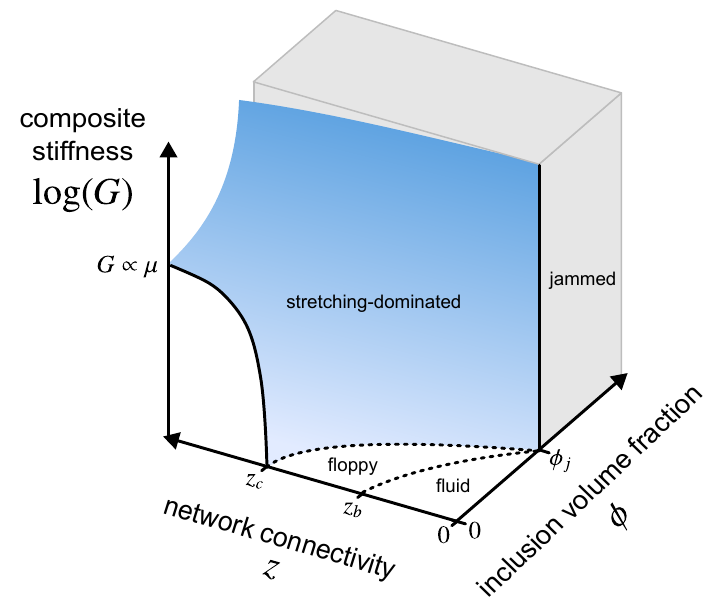}
    \caption{\label{fig:3D_schem_stretching_only}
      Schematic representation of the composite stiffness $G$ for disordered elastic networks with connectivity $z$ and rigid inclusion volume fraction $\phi$, for networks with exclusively central force stretching interactions. In the fluid regime, the composite does not form a space-spanning connected structure. In the floppy regime, the composite does form a space-spanning connected structure but is nonrigid ($G=0$). 
    }
  \end{figure}

In Fig. \ref{fig:3D_schem_stretching_only}, we present a schematic of the dependence of the composite stiffness $G$ on network connectivity $z$ and inclusion volume fraction $\phi$, for the central force (stretching only, i.e. $\kappa = 0$) limit.

\section{Scaling of $G$ with $\phi$ in the dilute regime}

\begin{figure*}[ht]
    \centering
    \includegraphics[width=0.8\textwidth]{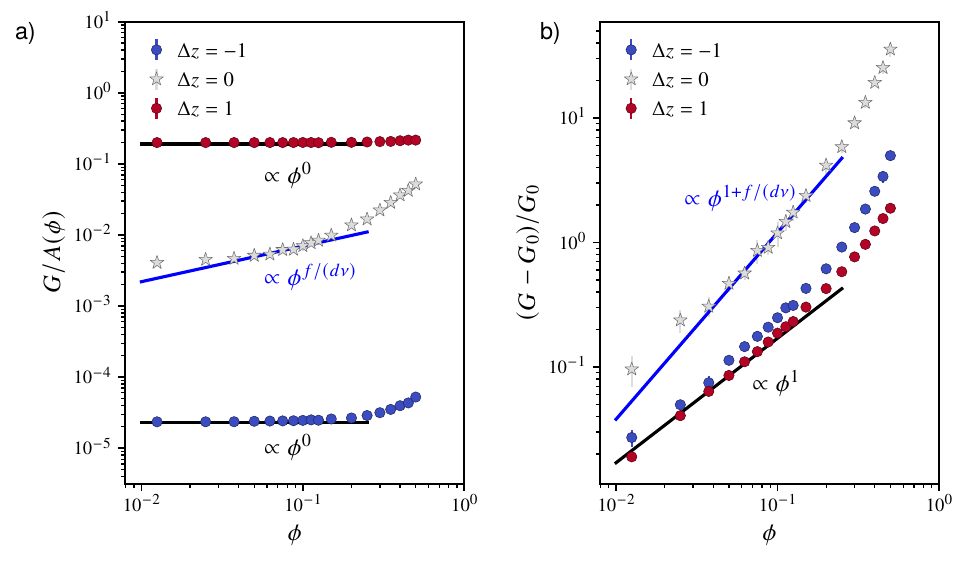}
    \caption{\label{fig:G_scaling_dilute}
     \textbf{The low-$\phi$ scaling predicted by the scaling theory is satisfied in an intermediate regime, failing in the extreme dilute limit}. (a) Shear modulus $G$ divided by stiffness amplification factor $A(\phi)$, as a function of particle volume fraction $\phi$. (b) Normalized excess shear modulus $(G-G_0)/G_0$ as a function of particle volume fraction $\phi$. Here $\kappa = 10^{-5}$ and each point represents an average over $n_s = 5$ samples. Error bars represent $\pm1$ standard deviation.
    }
  \end{figure*}

In Fig. \ref{fig:G_scaling_dilute}, we compare the dilute (low-$\phi$) scaling of $G$ observed in our simulations with the low-$\phi$ behavior predicted by the scaling theory. We find that our dilute-regime predictions are only satisfied in an intermediate-$\phi$ regime, failing in the extreme dilute regime. This suggests that the argument for treating the inter-inclusion distance $\xi_p$ as an effective finite network size is invalid in the extreme dilute regime, which one might expect: in the this limit, the particles should instead be treated as far-field defects in the network.

% \end{appendices}

% \clearpage
% \onecolumngrid

% \bibliography{Bibliography}{}


\end{document}